\documentclass[prb,twocolumn,aps,superscriptaddress,nofootinbib]{revtex4-2}

\usepackage[colorlinks = true,
            linkcolor = blue,
            urlcolor  = blue,
            citecolor = blue,
            anchorcolor = blue]{hyperref}
\usepackage{graphicx}% Include figure files
\usepackage{xcolor}
\usepackage{amsfonts}
\usepackage{amsmath}
\usepackage{mathtools}
\usepackage{float}
\usepackage{dcolumn}% Align table columns on decimal point
\usepackage{bm}% bold math
\usepackage{placeins}

\begin{document}

\preprint{APS/123-QED}

\title{On the microscopics of proximity effects in one-dimensional superconducting hybrid systems} 

\author{Siddhant Midha}
 \altaffiliation[Current Address: ]{Princeton Quantum Initiative, Princeton University, Princeton, NJ 08544, USA.}
 \affiliation{Department of Electrical Engineering, Indian Institute of Technology Bombay, Powai, Mumbai--400076, India
}\author{Roshni Singh}
 \altaffiliation[Current Address: ]{Department of Physics, University of California, Berkeley, CA 94720, USA.}
 \affiliation{Department of Physics, Indian Institute of Technology Bombay, Powai, Mumbai--400076, India
}\author{Kaveh Gharavi}
\affiliation{Institute for Quantum Computing, University of Waterloo, Canada}
\author{Jonathan Baugh}
\affiliation{Institute for Quantum Computing, University of Waterloo, Canada}
\author{Bhaskaran Muralidharan}
\email{bm@ee.iitb.ac.in}
 \affiliation{Department of Electrical Engineering, Indian Institute of Technology Bombay, Powai, Mumbai--400076, India
}
\affiliation{Center of Excellence in Quantum Information, Computing Science and Technology, Indian Institute of Technology Bombay, Powai, Mumbai--400076, India}

\date{\today}

\begin{abstract}
Investigating the microscopic details of the proximity effect is crucial for both key experimental applications and fundamental inquiries into nanoscale devices featuring superconducting elements. In this work, we develop a framework motivated by experiments to study induced superconducting correlations in hybrid nanoscale devices featuring layered superconductor-normal heterostructures using the Keldysh non-equilibrium Green's functions. Following a detailed method for analyzing the induced pair amplitude in a prototypical one-dimensional hybrid, we provide insights into the proximity effect within and outside the Andreev approximation. Our analysis also uncovers a disorder-induced crossover in the correlation patterns of the system. By elucidating the spectral distribution of the induced pair amplitude, we investigate the pair correlations established in a recent experiment \href{https://journals.aps.org/prl/abstract/10.1103/PhysRevLett.128.127701}{[Phys.Rev.Lett.128,127701]}, providing a theoretical basis for the enhanced Cooper pair injection demonstrated through the lens of the induced pair correlations, thereby establishing the promise of our methods in guiding new experiments in hybrid quantum devices. 

\end{abstract}
\maketitle

\section{Introduction}
Hybrid superconducting-semiconducting (super-semi) systems \cite{HybridSCSM_Review,HybridSCSM_QTech,HybridSCSM_dots} are a cornerstone in a multitude of current directions in quantum technology and condensed matter physics, enabling advancements in avenues such as the generation of spin-entangled electrons \cite{cps1,cps2,cps3,cps4,cps5,cps6,cps7,cps8}, Andreev qubits \cite{andreevbits0,andreev_bits1,andreev_bits2,andreev_bits3,andreevbits}, platforms for topological quantum computing \cite{topoqc,topoqc1,topoqc2,Majorana_CMP}, supercurrent transistors \cite{supertransistor1,supertransistor2,supertransistor3}, and superconducting quantum interference devices \cite{squid1,squid2,squid3,squid4,squid5}. The underlying physical phenomenon that allows many of these advances is the proximity effect, which can be explained by the Andreev reflection \cite{AndreevAndProximity,AndreevReflection_graphene,AndreevReflection_resonant,AndreevReflection_specular,AndreevReflection_theoryDOS,AndreevReflection_FM} at the superconducting-normal {(metal/semiconductor)} junction, whereby a spin-up electron incident at the junction on the normal side can undergo a retroreflection as a spin-down hole while creating a Cooper pair inside the superconductor. The retroreflected hole and the incident electron have a phase difference that depends on the superconducting gap. \\
\indent The proximity effect \cite{proximity_SC_TI,Proximity_SC_QD,Proximity_SCFM,Proximity_SC_Weyl,Proximity_SC_Graphene,Proximity_SCTI_,Proximity_SCTI__,Proximity_FMQDSC,Proximity_SC_FM_,Proximity_SCFM_Osc,Proximity_TripletSC,Proximity_SCFM_Review,Proximity_SCNM_HardGap,Proximity_TI_nanowires,Proximity_SC_TIGraphene,Proximity_SCFM_LongRange,Proximity_SC_FractionalTI,Proximity_SC_inhomogenous,Proximity_TI_SC_artifical,Proximity_SC_unconventional,Proximity_SC_theoryUnconventional,Proximity_SC_TunnelingModel,SFreview,ProxIndThin} is a tandem of such Andreev reflections at the junction and coherence properties of the normal conductor \cite{AndreevAndProximity}. Equivalently, the dual view, where a Cooper pair from the superconducting condensate leaks into the normal region, indicates a leakage of correlation effects from the superconductor into the non-ordered region. This in turn induces proximitized superconductivity in the normal region. Recent studies have explored the superconducting proximity effect on a variety of intriguing platforms, including graphene \cite{Proximity_SC_Graphene,Proximity_SC_TIGraphene}, topological insulators \cite{proximity_SC_TI,Proximity_SCTI_,Proximity_TI_supercurent,Proximity_SC_FractionalTI}, Weyl semimetals \cite{Proximity_SC_Weyl}, magnetic and ferromagnetic structures \cite{Proximity_SCFM,Proximity_FMQDSC,Proximity_SC_FM_,Proximity_SCFMhetero,Proximity_SCFM_LongRange}, and quantum dots \cite{Proximity_SC_QD,Proximity_FMQDSC}.\\
\indent Examining the nature of the pair correlations induced in normal (non-ordered) regions is thus of utmost importance in a way that allows for the understanding of the large body of devices fabricated for numerous applications, particularly those in which understanding the precise physics of the interface of a heterostructure is essential. Conventionally, the $s$-wave superconducting order is analyzed through the complex order parameter $\Delta$ as outlined in the microscopic BCS theory of superconductivity by Bardeen, Cooper, and Schrieffer \cite{bcs}. In heterostructures, one typically assumes a step function profile which is only non-zero in the superconducting region and sharply decays to zero in the normal region. However, induced correlations can be studied through variations in the amplitude \textit{of the pair} at a position $r$, given as $F(r) = \langle c^{\dagger}_{\uparrow r}c^{\dagger}_{\downarrow r}\rangle$, which is the correlator corresponding to the pairing process and is intuitively understood as the probability amplitude to find a Cooper pair at position $r$. The pair amplitude can show non-trivial decay patterns into the normal region, which reflects the nature of the proximitized correlations and thus serves as a probe to study induced superconducting correlations \cite{inducedqd,Proximity_SC_DisorderSelfcOnsistent,disorderpwave,Proximity_SC_Disorder,BdGSpinValve}.\\ 
\indent This approach can be applied more generally to a broader class of systems with interfaces between ordered and nonordered systems, such as ferromagnet-normal metal junctions, where one could study the spatial variation of the spin density difference in the normal region owing to a polarization present only in the ferromagnet. Moreover, in this work we limit ourselves to studying the induced $s$-wave order when in proximity to an $s$-wave superconductor. However, the approach outlined is general and can be straightforwardly adapted to different types of pairing as well. Important current research directions here are the effective induced $p$ wave superconductivity in Rashba nanowire super-semi hybrids \cite{sarma2015majorana,Proximity_SC_Disorder} and quantum dot-based minimal Kitaev chains \cite{minimalkitaevqd}.\\
\indent In this work, we set up a Keldysh non-equilibrium Green's function (NEGF) based approach to this problem, thereby outlining a general method to study correlations in hybrid quantum devices. Coupled with the existing capabilities of the NEGF method in describing quantum transport in experimentally relevant systems \cite{NEGF_transport_1,NEGF_transport_2,NEGF_transport_3,NEGF_transport_4,NEGF_transport_5}, we thus take an important step toward establishing a complete framework for engineering devices exploiting fermionic correlations in the solid state for quantum technology tasks. We illustrate this framework in simple one-dimensional superconducting-normal (SN) systems, where we calculate the pair amplitude in various regimes and study the interplay of the Andreev approximation in the pair correlations and density of states. We show how the expected behavior of the sub-gap conductance in the Andreev approximation is linked to the underlying proximity-induced correlations. \\ 
\indent We consider a bulk normal region to investigate the decay mechanisms of proximity-induced correlations. However, the results are also applicable to atomically thin normal regions, provided that suitable contacts for transport measurements are used. Further, by incorporating an experimentally motivated disorder potential at the junction, we identify a disorder-induced crossover in the system's correlation patterns. This, in turn, emphasizes the crucial need for a thorough understanding of the disorder mechanisms in order to effectively employ a device in experiments that depend on proximity-induced correlations. We also provide novel insight into the spectral distribution of the pair amplitude across the device and study a recent proposal for a resonant Cooper pair injection device through the lens of induced correlations. This study focuses primarily on one-dimensional systems; however, the methods developed are quite general and can be extended to account for higher-dimensional effects, such as transverse sub-bands, using the approach outlined in \cite{praveen_}.\\  

\indent This paper is organized as follows. In Sec.~\ref{sec:model}, we define the lattice model under study and outline the NEGF method. We then begin with the results in Sec.~\ref{sec:SNclean} for clean nanowires with superconductor-normal (SN) interfaces and then study the effects of the Andreev approximation in Sec.~\ref{sec:andreev}. Then, the spectral distribution of the pair amplitude is illustrated in Sec.~\ref{sec:spectral}, and a disorder-induced crossover in the correlation patterns is shown in Sec.~\ref{sec:disorder}. The study of the Cooper pair injection device is then described in Sect.~\ref{sec:cpinj}. Finally, we conclude the paper and discuss future work in Sec.~\ref{sec:conc}.

\section{The NEGF Model} \label{sec:model}
\begin{figure}[t]
\includegraphics[width=0.45\textwidth, height = 10cm]{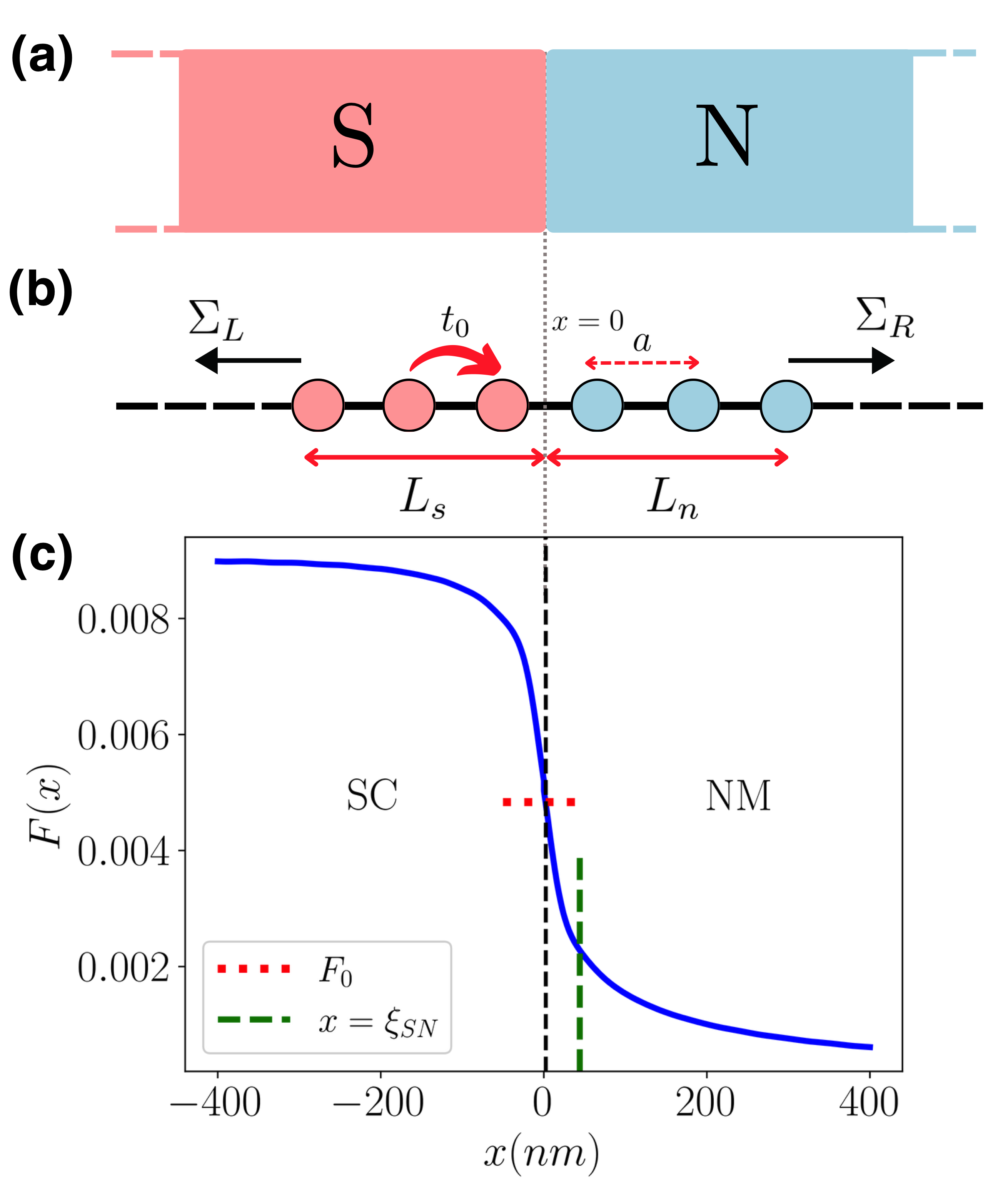}
\caption{\textbf{Pairing amplitude profile across the SN junction in a hybrid nanowire setup}. (a) Schematic of the device showing the effective one-dimensional nanowire with the SN junction at $x=0$. (b) One-dimensional tight-binding model of the nanowire within the NEGF framework, showing the lattice discretization $a$, hopping strength $t_0$, contact self energies $\Sigma_{S/N}$, and the number of lattice sites considered in the Hamiltonian $L_{s/n}$. (c) Pair amplitude across the SN junction exhibiting the non-trivial decay of superconducting correlations inside the normal region. $\Delta=1$meV and $\mu = 10$meV. The pair amplitude at the interface $F_0 \equiv F(x=0)$, and the effective coherence length $\xi_{SN} = 44(1)$nm for the SN junction are indicated on the plot.} \label{fig:figclean}
\centering
\end{figure}

We consider a one-dimensional hybrid SN nanowire setup, as shown in Fig.~\ref{fig:figclean}(a), as the prototypical system for illustrating the physics of the proximity effect. The junction is located at $x=0$ (dotted vertical line in Fig.~\ref{fig:figclean}(a)), the normal side is located at $x>0$ and the superconducting side at $x<0$. A lattice discretization of this nanowire is shown in Fig.~\ref{fig:figclean}(b), with the discretization of $a=2$nm. A non-interacting Hamiltonian consisting of a spatially modulated pair potential describing the wire reads,
\begin{multline}\label{eq:hamil}
    \mathcal{H} = (2t_0-\mu)\sum_{i,\sigma}c_{i\sigma}^{\dagger}c_{i\sigma} - t_0\sum_{\langle ij\rangle,\sigma}(c_{i\sigma}^{\dagger}c_{j\sigma} + h.c.) \\ 
    + \Delta(x)\sum_i (c_{i\uparrow}^{\dagger}c_{i\downarrow}^{\dagger} + h.c.)
\end{multline}
where $h.c.$ denotes the Hermitian conjugate, $\mu$ denotes the chemical potential of the system, $t_0=\hbar^2/(2m_*a^2)$ denotes the hopping potential on the lattice, $\langle ij\rangle$ denotes nearest-neighbour interactions, and $c^{\dagger}_{i,\alpha}(c_{i,\alpha})$ creates(destroys) a $\alpha$-spin electron at site $i$ on the lattice. Furthermore, $\Delta(x)$ follows a step-function profile $\Delta(x) = \Delta \Theta(-x)$ with $\Theta(x)$ being the Heaviside function. We take $\Delta$ to be a real constant owing to the presence of a single superconductor. Thus, the pair potential is nonzero only in the superconducting region of $x < 0$.\\ 
\indent The NEGF method works by specifying a finite segment of the nanowire as the system and accounting for the transport through the rest of the infinite wire by treating it as a bath. The bath's surface Green's functions are recursively computed in a self-consistent manner, which are then used to compute the self-energies of the left and right contacts in order to evaluate the system's Green's functions. The correlations present in the system along with the currents through the nanowire can then be evaluated through the Green functions. We take the system to be composed of $L_s = 200$ superconducting sites and $L_n = 200$ normal sites. If we denote the surface Green's function of the left(right) contact as $g_{L(R)}(E)$, and the tunnel coupling between the system and the left(right) contact as $\tau_{L(R)}$, then the (non-Hermitian) self-energies as a function of energy are given as,
\begin{equation}
    \Sigma_{L(R)}(E) = \tau_{L(R)}g_{L(R)}(E)\tau^{\dagger}_{L(R)},
\end{equation}
with the self-energy functions at hand, we define the contact broadening matrices,
\begin{equation}
    \Gamma_{L(R)}(E) =\iota\left[\Sigma_{L(R)}(E) - \Sigma^{\dagger}_{L(R)}(E)\right],
\end{equation}
which describe the level-broadening of the system upon coupling to the contacts. These are nothing but the non-Hermitian part of the self-energy, resulting in a finite lifetime of energy eigenstate and hence broadening the discrete spectrum into a continuum density of states. Now, we define the system's energy-resolved retarded Green's function $G^r(E)$ as
\begin{equation}
    G^r(E) = \left[(E+\iota\eta) - \mathcal{H} - \Sigma_L(E) - \Sigma_R(E)\right]^{-1},
\end{equation}
and the advanced Green's function as $G^a(E) = [G^r(E)]^{\dagger}$. The spectral function, given as,
\begin{equation}
    A(E) = \iota\left[G^r(E) - G^a(E)\right],
\end{equation}
is used to evaluate the local density of states through the diagonal elements,
\begin{equation}
    \text{LDOS}(x,E) = \frac{1}{2\pi}A(x,x;E),
\end{equation}
and the total density of states as follows,
\begin{equation}
    D(E) =\frac{1}{2\pi}\text{Tr}[A(E)].
\end{equation}
Furthermore, given the bath potentials $\mu_{L(R)}$ and the temperatures $T_{L(R)}$, the in-scattering functions can be defined by weighing the broadening matrices with the equilibrium Fermi distributions, 
\begin{equation}
    \Sigma_{L(R)}^{in}(E) = f(E,\mu_{L(R)},T_{L(R)})\cdot \Gamma_{L(R)}(E)
\end{equation}
which can then be used to define the current operators $I_{L(R)} = \Sigma_{L(R)}^{in}A - \Gamma_{L(R)}G^r$ used to compute the local and overall current density upon suitable integration over energy.

The induced proximity effect in the system can be examined via defining the correlation Green's function $G^n$,
\begin{equation}
    G^n(E) = G^r(E)[\Sigma_L^{in}(E) + \Sigma_R^{in}(E)]G^a(E)
\end{equation}
which encodes the spectral distribution of the two-point fermionic correlators. More specifically, if we define the Nambu spinor $\hat{\psi}_i^{\dagger} := (c_{i\uparrow}^{\dagger},c_{i\downarrow})$, then, 
\begin{equation}\label{eq:corrs}
    -\iota\langle\hat{\psi}_i^{\dagger}\hat{\psi}_j \rangle = \frac{1}{2\pi}\int [G^{n}(E)]_{ij}dE
\end{equation}
where $[M]_{ij}$ denotes a sub-block corresponding the $i^{th}$ row and $j^{th}$ column of the matrix $M$, which in the case for $M = G^n$ is a $2\times 2$ matrix.\\
\indent The superconducting order parameter $\Delta$ arises from a mean-field treatment of a four-term attractive interaction, leading to a Hamiltonian described by Eq.~\ref{eq:hamil}. This \textit{pair potential} $\Delta(x)$ is thus evaluated to a product of the interaction potential $g(x)$ (of the four-term interaction) and the \textit{pair amplitude} $F(x)$, $\Delta(x) \equiv F(x)g(x)$. Since the interactions are present only in the superconducting region, $g(x)$ follows a step-function profile, implying also the same behavior for $\Delta(x)$. However, $F(x) \equiv \langle c_{x\uparrow}^{\dagger}c_{x\downarrow }^{\dagger}\rangle$ need not exhibit such behavior, and instead shows a decaying profile in the normal region. This is a diagnostic of the pair correlations induced inside the normal region owing to proximity with the superconductor. Gaining insight into the nature of the decay of correlations into the normal region is thus crucial to successfully design devices with such a hybrid junction for various applications exploiting induced superconductivity. We now employ the NEGF method discussed here to systematically describe the microscopics of the proximity effect.

\begin{figure*}[t]
\includegraphics[width=0.8\textwidth, height = 10cm]{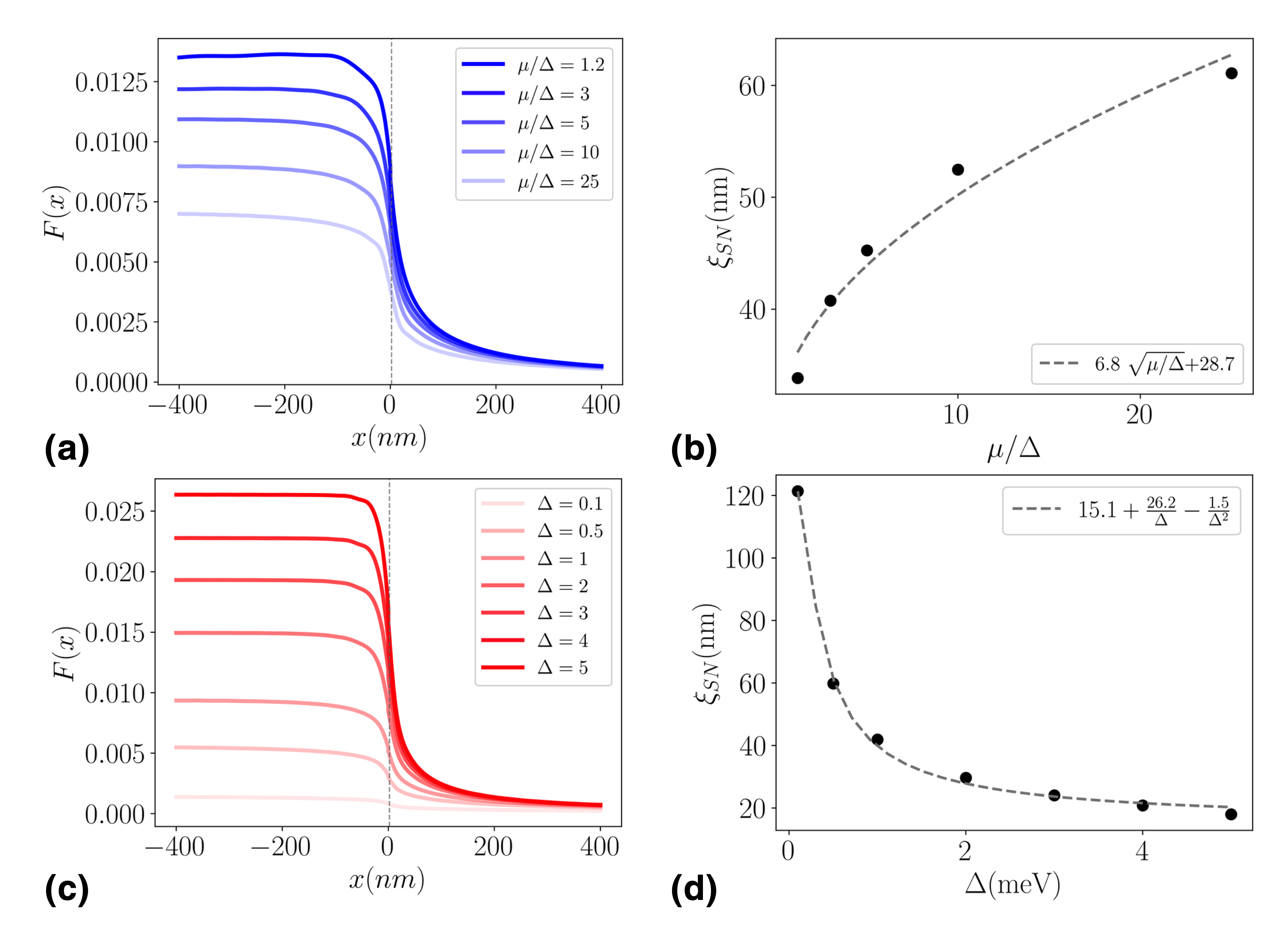}
\caption{\textbf{In and out of the Andreev approximation}. Top row (a-b): The chemical potential is varied in the range $\mu/\Delta \in \{1.2,3,5,10,15\}$ with a constant superconducting order $\Delta = 1$meV. (a) Pair amplitude computed across the SN junction for different values of $\mu/\Delta$. (b) Effective coherence length plotted as a function of $\mu/\Delta$ with a fitted curve of $\xi_{SN}(\mu/\Delta) = 6.8\sqrt{\mu/\Delta} + 28.7$(nm) superposed. Bottom row (c-d): The superconducting order parameter is varied in the range of $\Delta \in \{0.1,0.5,1,2,3,4,5\} $meV for a constant chemical potential $\mu = 10$meV. (d) Pair amplitude computed across the SN junction for different values of $\Delta$. (e) Effective coherence length as a function of $\Delta$ with a fitted curve of $15.1 + 26.2/\Delta - 1.5/\Delta^2~$(nm).} \label{fig:figandreev}
\centering
\end{figure*}

\begin{figure*}[t]
\includegraphics[width=1.0\textwidth, height = 9cm]{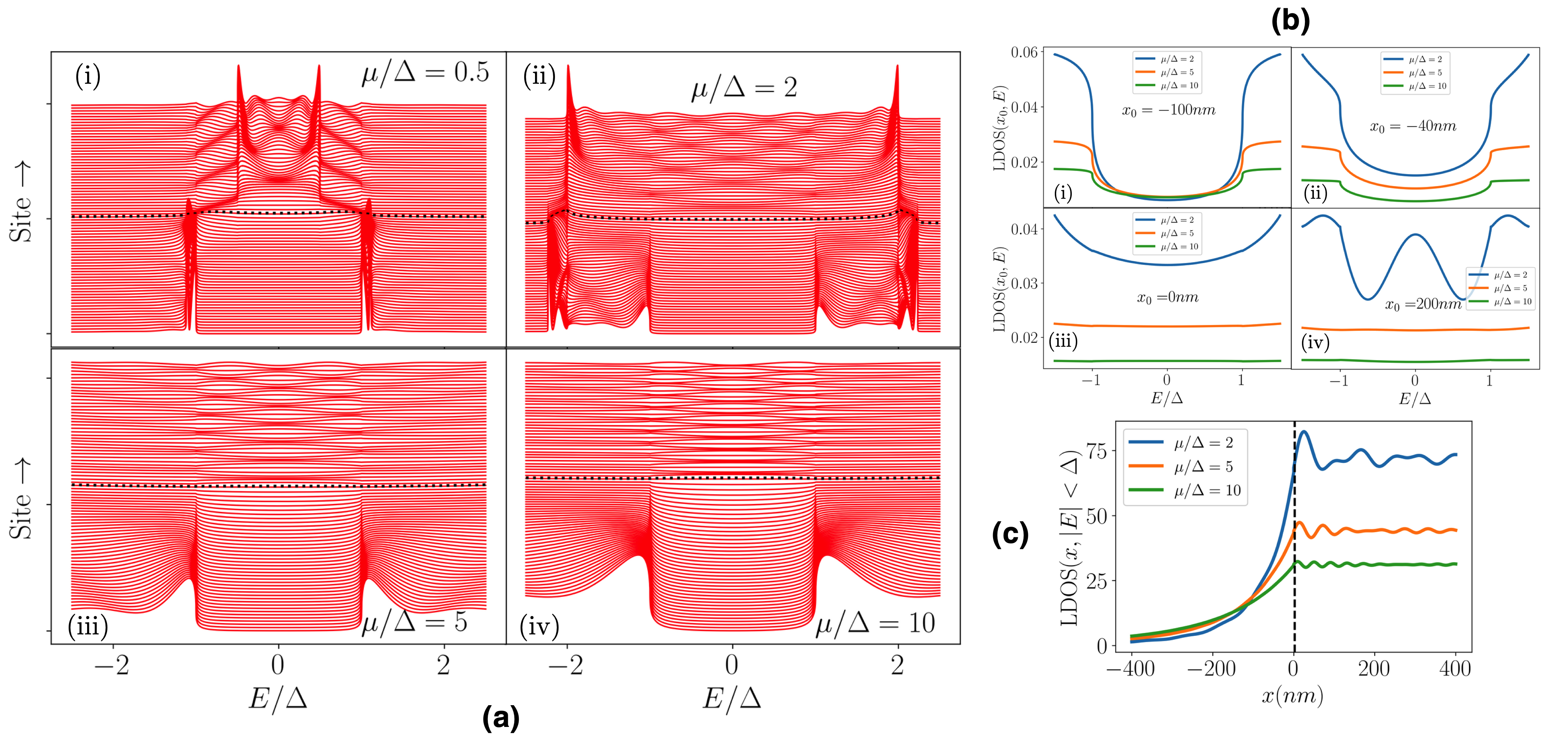}
\caption{\textbf{Density of states across the SN junction} (a) Local-density of states across the SN junction as a function of energy ($x-$axis) and lattice site (shifted vertically) with the dotted line marking the interface boundary separating the superconducting part (below the boundary) and the normal part (above the boundary) for a fixed order parameter $\Delta = 1$meV and chemical potentials $\mu/\Delta \in \{0.5,2,5,10\}$. (b) Energy resolved local density of states LDOS$(x_0,E)$ at fixed sites $x_0$ across the junction: (i) in the superconducting segment for $x_0 = -100$nm and (ii) $x_0 = -40$nm, at the interface $x_0 = 0$, and in the normal region $x_0 = 200$nm for $\mu/\Delta\in\{2,5,10\}$. (c) Total sub-gap local density of states across the nanowire obtained by integrating the LDOS in the $|E| < \Delta$ range as a function of $x$ across the junction for the same values of chemical potential as in (b).} \label{fig:figldos}
\centering
\end{figure*}

\begin{figure*}[t]
\includegraphics[width=0.8\textwidth, height = 10cm]{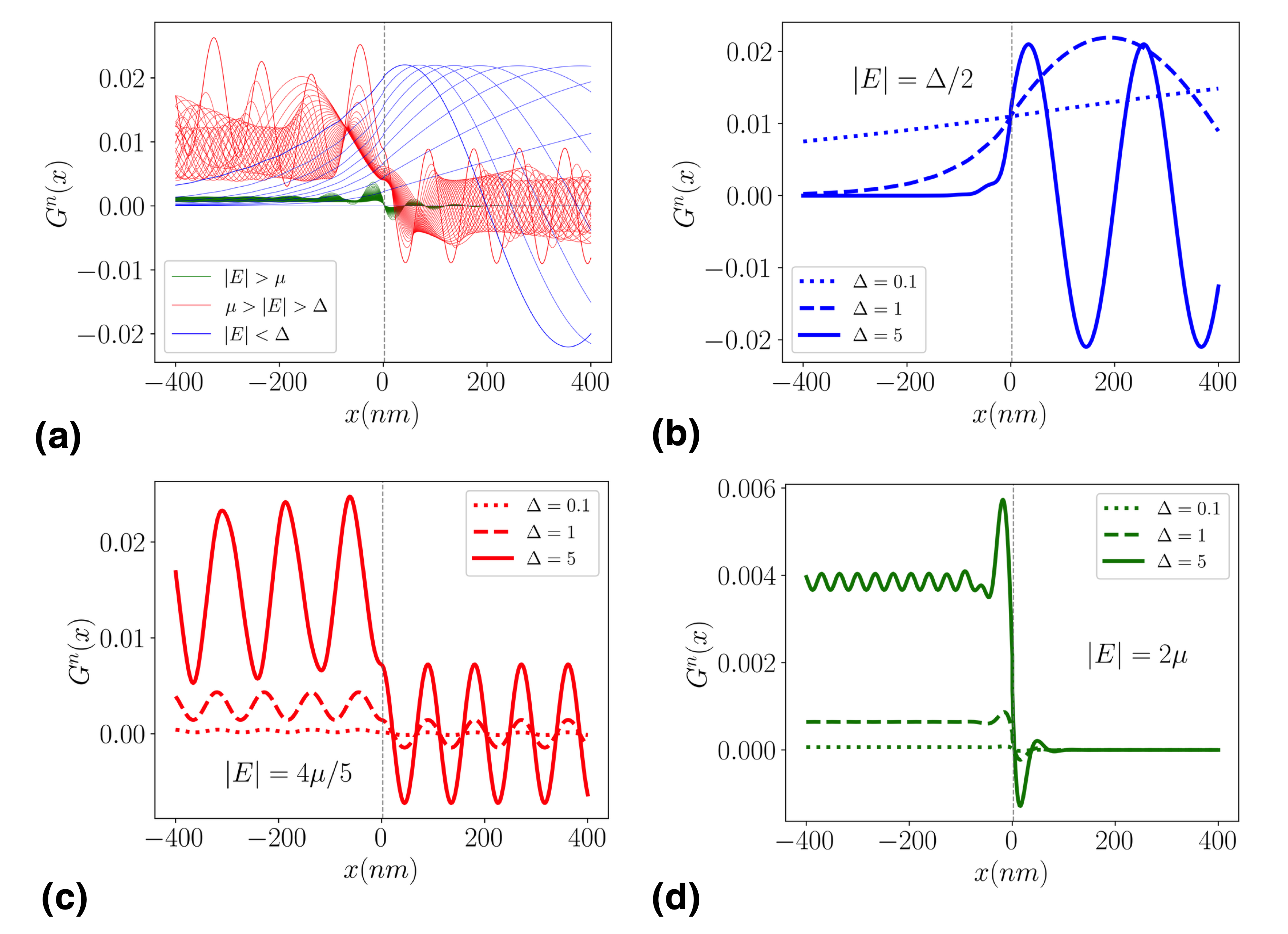}
\caption{\textbf{Spectral resolution of the pair amplitude}. (a) The correlation Green's function $G^n(x;E)$ plotted across the interface $x \in [-400,400]~\text{nm}$ for the three energy regimes $|E| < \Delta $, $\Delta < |E| < \mu$, and $|E| > \mu$ for $\Delta = 1$meV and $\mu = 10$meV. (b-d) Plot of $G^n(x;E)$ for three different order parameters $\Delta \in \{0.1,1,5\}~$meV and $\mu = 10$meV at specific points in the three energy regimes: (b) inside the gap at $|E| = \Delta/2$, (c) outside the gap at $|E| = 4\mu /5$, and (d) far outside the gap at $|E| = 2\mu$.} \label{fig:figspectral}
\centering
\end{figure*}

\begin{figure*}[t]
\includegraphics[width=1.0\textwidth, height = 8cm]{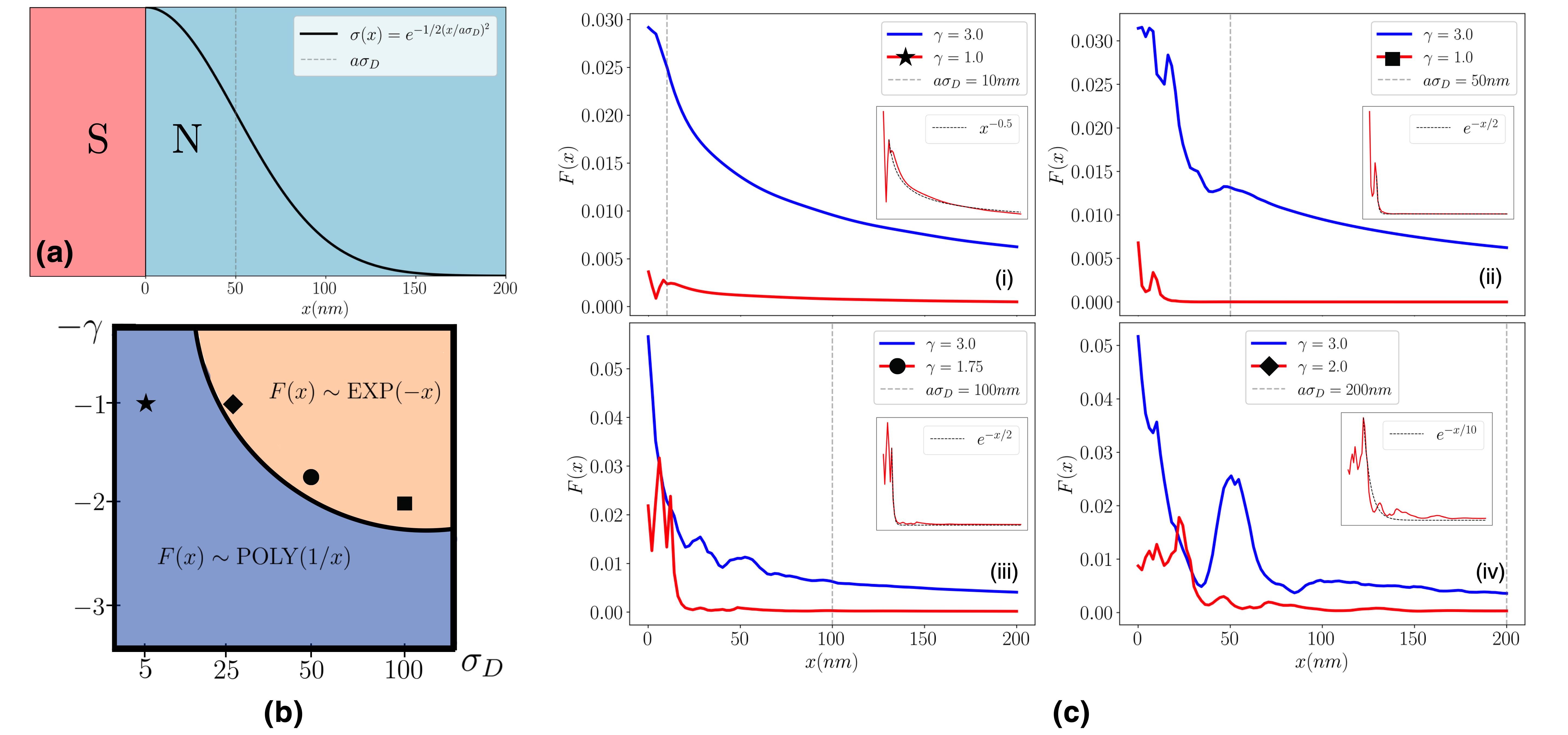}
\caption{\textbf{Disorder induced crossover} (a) Schematic of the disorder profile in the normal region, with $\sigma(x)$ denoting the variance of the zero-mean uncorrelated potential at each location $x\geq 0$, and the spread coefficient $\sigma_D$ denoting the spatial extent of $\sigma(x)$ into the normal region (b) A map of the induced correlations profile under disorder as a function of $\gamma$ (decay coefficient specifying the order of the potential $V/t_0 \sim 10^{-\gamma}$) and $\sigma_D$ (spread of variance profile) consisting of four data points and a sketch of the crossover line separating the two decay behavior (c) Pair amplitude for (i) $\sigma_D=5,~\gamma \in \{3,1\}$ (ii) $\sigma_D=25,~\gamma \in \{3,1\}$, (iii) $\sigma_D=50,~\gamma \in \{3,1.75\}$ and (iv) $\sigma_D=100,~\gamma \in \{3,2\}$ with the red curves shown in the inset with polynomial/exponential fits to the tails and corresponding to the marked points on the schematic. The vertical dotted line denotes the variance of the disorder profile $a\sigma_D$.} \label{fig:figdisorder}
\centering
\end{figure*}

\begin{figure*}[t]
\includegraphics[width=\textwidth, height = 8cm]{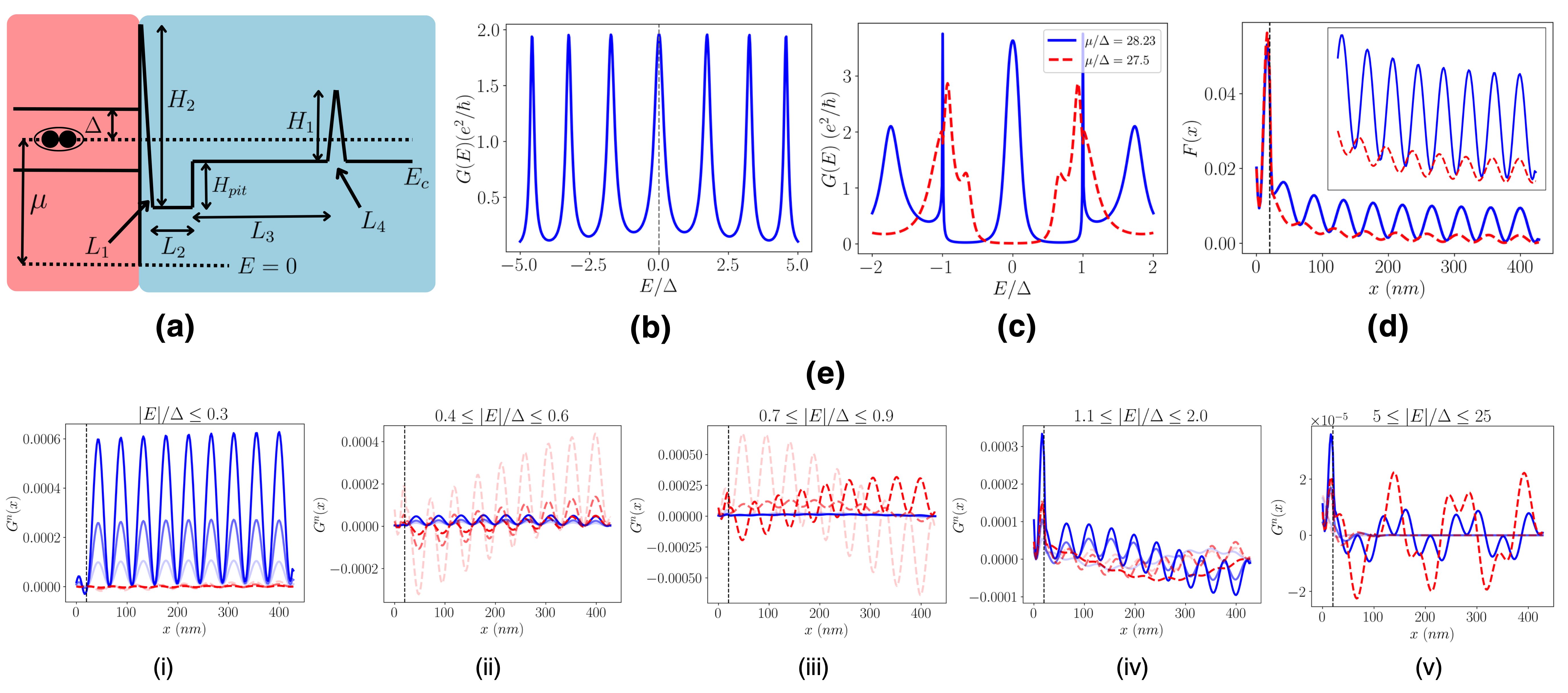}
\caption{\textbf{Analyzing resonant Cooper-pair injection via the pair amplitude}. (a) Schematic toy potential model of the resonant device in \cite{CPInjection_ExpMain} showing a superconducting segment in proximity to a semiconductor quantum well resonant state. Parameters are $H_2 = 180\text{meV}, H_{pit} = 100\text{meV}, H_1 = 160\text{meV}, \mu - E_c = 20\text{meV}$ and $L_1 = 2\text{nm}, L_2 = 4\text{nm}, L_3 = 400\text{nm}, L_4 =2\text{nm}$ with $t_0 = 415\text{meV}$. (b) Transmission spectrum of the quantum well (without proximity to a superconductor). (c) Transmission spectrum of the proximity-coupled device for the resonant case of $\mu_{1} = 28.23~\text{meV}$ and the non-resonant case of $\mu_{2} = 27.5~\text{meV}$ and $\Delta=1~\text{meV}$. (d) Pair amplitude profile across the device for $\mu/\Delta \in \{28.23,27.5\}$, with the superconducting-normal interface marked with a dotted line. Inset shows a zoomed-in plot of the pair amplitude only in the quantum well region. (e) Energy resolved pair amplitude profile: The correlation Green's function is plotted for (i) $|E|/\Delta = \{0.1,0.2,0.3\}$, (ii) $|E|/\Delta = \{0.4,0.5,0.6\}$, (iii) $|E|/\Delta = \{0.7,0.8,0.9\}$, (iv) $1.1 \leq |E|/\Delta = \{1.1,1.5,2.0\}$, (v) $|E|/\Delta =\{5,10,25\}$ with the plot transparency decreasing with increasing $|E|/\Delta$ within each subplot.} \label{fig:figcpinj}
\centering
\end{figure*}

\section{Results and Discussion}\label{sec:results}
\subsection{Induced superconducting correlations in SN junctions}\label{sec:SNclean}
We now discuss the behavior of the pair amplitude evaluated using the NEGF method discussed in the previous section for the prototypical SN junction of Fig.~\ref{fig:figclean}(a). The result is shown in Fig.~\ref{fig:figclean}(c) for the case where the chemical potential is $\mu = 10$meV and the order parameter is $\Delta = 1$meV, thereby satisfying the Andreev approximation of $\mu/\Delta \gg 1$. We consider the system at zero temperature, and with no voltage/thermal drive on the baths. Deep in the superconducting side $x \to -\infty$, the pair amplitude attains a constant non-zero value signifying the bulk correlations in the superconductor. As we move closer to the interface $x \to 0$, the amplitude shows a decaying profile into the normal region, and decays to zero deep in the normal region $x \to \infty$. We denote the value of $F(x)$ at the junction as $F_0 \equiv F(x=0)$. The decay of the pair amplitude into the normal region is found to be power law of the form,
\begin{equation} \label{eq:powerlaw}
    F(x) = F_0 \frac{\xi_{SN}}{x + \xi_{SN}}~(x\geq 0)
\end{equation}
where $\xi_{SN}$ is thus interpreted as an effective coherence length of the junction, defining a length-scale for algebraically decaying correlations ($F(x)\to F_0/2 \text{ as } x\to \xi_{SN}$). In Fig.~\ref{fig:figclean} we find $\xi_{SN} = 44(1)$nm. Moving further, the effective junction coherence length is obtained by fitting to the form of Eq.~\ref{eq:powerlaw} unless otherwise specified. The power-law decay of the pair amplitude, in turn, indicates a long-range penetration of the induced correlations into the normal region. We attribute this to the absence of disorder in the normal region, which leads to an infinite coherence length, since the retroreflected particle does not lose the phase information regarding the superconductor by scattering events in the non-ordered region. This finding is in agreement with the power-law order parameter decay found in the case of clean nanowires at zero temperature in Ref.~\cite{Proximity_SC_DisorderSelfcOnsistent} where a self-consistent Bogoliubov-de Gennes (BdG) approach is employed. In doing so, this establishes the correctness of our method thereby benchmarking with traditional self-consistent calculations.\\
\indent We take a moment here to stress here the importance of self-consistency: it is imperative to have a self-consistent theory which results in the correct non-trivial decay of the pair amplitude upon the input of a heavyside-like pair potential in the mean-field picture. The NEGF method implicitly takes this into account through the self-consistent calculation of the Green's functions of the bath and subsequently of the system. This offers insights not only into the decay of induced superconducting order in the normal region but also into the backaction of the normal region on the superconductor. Consequently, it enables a comprehensive understanding of the system's behavior both near the interface and deep within the bulk. Furthermore, the NEGF method sets out to be more general than the self-consistent BdG methods owing to the fact that the system is not required to be in specific boundary geometries, and one can naturally describe correlations in setups used for quantum transport measurements via attaching source and drain contacts to the system. That is, instead of isotropic case studied here (wherein $\tau_L=\tau_R=t_0$, making the system translation invariant on both sides of the origin) one can consider weakly coupled probes with $\tau_{L/R} < t_0$ to investigate realistic transport measurements.  

\subsection{Effect of the Andreev approximation}\label{sec:andreev}
Approaches based on scattering theory typically rely on the system being in the Andreev approximation $\mu \gg \Delta$ \cite{andreev}, which ensures that only Andreev reflections (and thus no normal reflections) happen at the superconducting-normal interface. This assumption is not a prerequisite when using the NEGF method. We thus proceed to utilize the methods developed to study the interplay of the $\mu/\Delta$ parameters on the proximity effect. It is well known that as we go deeper into the Andreev approximation with increasing $\mu/\Delta$, the sub-gap conductance of a clean NS nanowire shows a pristine $4e^2/h$ conductance peak instead goes to zero in the opposite $\mu/\Delta \to 0$ limit \cite{analytic_andreev}. We will demonstrate that the proximity-induced correlations exhibit unexpected behavior, yet ultimately align with the established conductance picture. Furthermore, we will explore how the coherence properties of the hybrid nanowire evolve as superconducting order increases.\\ 

\indent The results for the pair amplitude where we tune in and out of the Andreev approximation are shown in the top row of Fig.~\ref{fig:figandreev}. We see that, when for a fixed $\Delta=1$meV as we increase the ratio $\mu/\Delta$ in the set $\{1.2,3,5,10,25\}$, the bulk value of pair amplitude inside the superconductor \textit{decreases}, as the in-gap excitations are decreased along with an increase in the chemical potential. This further implies that the induced amplitude inside the normal region {decreases} upon tuning into the Andreev approximation. But, upon careful inspection, it can be shown that the effective coherence length of the junction $\xi_{SN}$ increases with the value of $\mu/\Delta$. This is shown in Fig.~\ref{fig:figandreev}(b), where the growth of the effective coherence length is $\sim \sqrt{\mu/\Delta}$. This is extracted by fitting $\sim F_0\xi_{SN}/x$ to the tail of the pair amplitude curves at $x \gtrsim 200$nm, as we expect the coherence length away from the interface to be related to the only length-scale present in the ballistic system---the Fermi wavelength, which depends on $\sqrt{\mu}$---upto finite size effects. The numerical coefficients are phenomenological, and expected to change with the microscopic parameters. Moreover, this increase in coherence implies that the pair correlations leaking into the normal region increase as we go deeper into the Andreev approximation regime, which is in agreement with the sub-gap conductance picture.\\
\indent This is also seen by examining the density of states in the system, which is shown in Fig.~\ref{fig:figldos}. Shown in Fig.~\ref{fig:figldos}(a) is the local density of states (LDOS) across the system with energy on the $x$-axis and position on the $y-$axis, with a horizontal dotted line marking the junction. The signature superconducting gap in the $-\Delta \leq E \leq \Delta$ region is visible, interacting with the in-gap excitations of the normal region for the different cases of $\mu/\Delta \in \{0.5,2,5,10\}$. As we move from $\mu/\Delta=0.5$ to $\mu/\Delta = 10$, the absence of in-gap excitations in the normal region is clearly seen. To further illustrate the change in density of states as $\mu/\Delta$ changes, Fig.~\ref{fig:figldos}(b) displays the local density of states at specific site locations, namely inside the superconductor at $x_0 = -100\text{nm}$ and $x_0 = -40\text{nm}$, at the junction $x_0 = 0$, and inside the normal region $x_0 = 200\text{nm}$ for $\mu/\Delta\in \{2,5,10\}$. Deep inside the superconductor at $x_0 = -100\text{nm}$, the in-gap density is essentially the same for the three cases. An \textit{inverse} proximity effect--the back-action of the normal side on the superconductor--can be seen at $x_0=-40\text{nm}$, where the in-gap density in the superconductor decreases as we move deeper in the Andreev approximation. These inverse effects are particularly relevant in superconducting-ferromagnetic systems \cite{SFreview}, and are not discussed further here. The LDOS at the junction $x_0=0$ shows a similar behavior as $\mu/\Delta$ increases. Further, the in-gap excitations at low $\mu/\Delta=2$ can be seen in the normal region at $x_0=200\text{nm}$, which dampens out deeper in the Andreev approximation regime for $\mu/\Delta=5,10$. A summary of these effects is further shown as the total integrated in-gap density of states [$\int_{|E| < \Delta} \text{LDOS}(x,E)dE$] across the junction in Fig.~\ref{fig:figldos}(c).\\ 
\indent Finally, we outline how the effective coherence length of the junction evolves as the superconducting order is increased. This is shown in the bottom row of Fig.~\ref{fig:figandreev}. We plot the induced pair amplitude for a fixed chemical potential of $\mu=10$meV for the different order parameter values $\Delta \in \{0.1,0.5,1,2,3,4,5\}$meV in Fig.~\ref{fig:figandreev}(c). We see an increasing value of the pair amplitude in the bulk superconductor and the induced amplitude in the normal region as $\Delta$ is increased. The effective coherence lengths are then plotted in Fig.~\ref{fig:figandreev}(d). We note an algebraic decay in $1/\Delta$ of the effective coherence length, which goes as $\sim 1/\Delta$ for large $\Delta$. This highlights an important consideration when designing hybrid devices that leverage the proximity effect: two different values of the nearby superconducting order, $\Delta_1$ and $\Delta_2$, can yield nearly identical sub-gap conductance in the near-pristine limit (e.g., when $\mu/\Delta_1, \mu/\Delta_2 \gg 1$, as in the case of $\Delta_1 = 0.1$ meV and $\Delta_2 = 0.5$ meV here). However, they will induce superconducting order that propagates differently within the normal region, as reflected in the effective coherence length. Moreover, although this algebraic dependence resembles the coherence length of a BCS superconductor ($\xi_S \sim 1/\Delta$), it must be noted that the two quantities describe fundamentally different phenomena. The BCS coherence length represents the average size of a Cooper pair within the superconductor, whereas the effective coherence length of the junction here quantifies the spatial extent of the induced correlations within the normal region.

\subsection{Spectral resolution of the pair amplitude}\label{sec:spectral}
We have highlighted that analyzing the nature of the induced pair amplitude uncovers valuable information that is not readily accessible through conventional methods, such as transport measurements. Next, we will explore how this method can be used to derive more fundamental insights into the induced pair correlations in generic hybrid systems by outlining its energy-resolved structure. The pair amplitude $F(x)$, which can be interpreted as the probability amplitude of finding a Cooper pair at position $x$, is a very useful diagnostic of such correlations and can also be calculated using self-consistent BdG-like approaches (for instance, see \cite{Proximity_SC_DisorderSelfcOnsistent,BdG2D,BdGquasic,BdGSpinValve}) albeit in restricted geometries and without a connection to practical hybrid devices. Our approach however allows the computation of the pair amplitude, in addition to providing insight into its spectral distribution and thus potentially engineer fabricated devices to optimize quantities of interest. \\

\indent The spectral distribution of the pair amplitude for the prototypical clean SN junction is shown in Fig.~\ref{fig:figspectral}. The spectrally distributed correlations across the length of the nanowire for the three regimes of interest, namely $|E| < \Delta$, $\Delta <|E| < \mu$ and $|E| > \mu$ are shown in Fig.~\ref{fig:figspectral}(a). These individual traces average out to give the net pair amplitude as previously shown in Fig.~\ref{fig:figclean}(c). Starting with the $|E| < \Delta$ case, we see an \textit{increasing} trend moving from the superconductor to the normal region, which ultimately oscillates with small frequencies. Thus, this is the region in energy space where the correlations from the superconductor couple to the normal region leading to proximity-induced correlations. Moving further, we see the expected step-like trend in the $\Delta <|E| < \mu$ region, where the correlation curves oscillate at various frequencies. It is an interplay of these oscillations which add up in the integral of Eq.~\ref{eq:corrs} and interfere to produce a net pair amplitude, which could be power-law decaying in the case of clean nanowires and show exponential decay in disordered wires. The nature of decay is thus obtained by the interference patterns of the individual spectrally distributed correlations. Lastly, we see that far-off energies $|E| > \mu$ have negligibly small contributions to the net pair amplitude, which allows for a truncation of the integral in Eq.~\ref{eq:corrs} in practice.\\ 

\indent To further illustrate these points, we show three particular energy distributed curves, for the three cases of $\Delta\in\{0.1,1,5\}$meV. Fig.~\ref{fig:figspectral}(b) shows the results for the in-gap energy of $|E|=\Delta/2$, Fig.~\ref{fig:figspectral}(c) shows the results for $|E|=4\mu/5$, and the far-off case of $|E|=2\mu$ is shown in Fig.~\ref{fig:figspectral}(d). For the in-gap case, we see that the frequency of oscillations inside the normal region increases as $\Delta$ is increased, which intuitively results in more destructive interference leading to a reduction in the proximity effect, in agreement with our previous results. But, for the $|E|=4\mu/5$ case, we also see that the magnitude of the normal region correlations also increases with the increase in $\Delta$, which is expected as the net superconducting order increases with $\Delta$, even if the coherence is decreased. Moreover, we note that the frequency of the oscillations in the $|E|=\Delta/2$ (Fig.~\ref{fig:figspectral}(b)) depend on the order parameter $\Delta$, and instead that of $|E|=4\mu/5$ (Fig.~\ref{fig:figspectral}(c)) depend only on the chemical potential. Finally, we see almost step-like curves decaying to zero nearly instantaneously in the normal region for the far-off energy case of $|E|=2\mu$.

\subsection{Disorder in the SN junction}\label{sec:disorder}
Anderson's theorem for bulk superconductors \cite{anderson1959theory} does not apply to proximity-induced pair amplitudes, making it imperative to study the effects of non-magnetic disorder on the proximity effect. Moreover, understanding how disorder influences proximity effects is crucial for utilizing the proximity effect in real-world devices. Previous work \cite{disorderpwave,Proximity_SC_DisorderSelfcOnsistent} explored the influence of random non-magnetic disorder on proximity-induced correlations, demonstrating that the induced $s$-wave amplitude decreases as disorder increases. Taking a step further, we consider an experimentally motivated disorder profile and provide a more complete characterization of the decay patterns as a function of the strength and extent of the disorder. This analysis reveals a crossover from power law to exponentially decaying correlations in the normal region. This is consistent with the crossover from weak to strong disorder observed in \cite{Proximity_SC_DisorderSelfcOnsistent}. While finite size scaling remains difficult owing to the extensive nature of the numerical computations, we conjecture that this crossover marks an order-disorder phase transition in the steady state of the hybrid system at zero temperature.   \\
\indent We examine the same one-dimensional system, but introduce a simple disorder profile on the normal side of the junction. Specifically, we impose a spatially varying variance amplitude, $\sigma(x)$ ($x \geq 0$), and model the disorder using independent, identically distributed Gaussian random potentials at each site\footnote{$X\sim \mathcal{N}(\mu,\sigma)$ denotes a Gaussian distributed random variable $X$ with $\mathbb{E}[X]=\mu$ and $\mathbb{E}[X^2] - \mathbb{E}[X]^2 = \sigma^2$}, 
\begin{equation}
    V(x)/t_0 \sim \mathcal{N}[\mu = 0, \sigma =\sigma(x)] \times 10^{-\gamma}
\end{equation}
such that the variance is a function set by $\sigma(x)$, and $\gamma$ sets the scale of the random potentials by ensuring $\sigma(x) \leq 1$ for all $x \geq 0$. We consider the values $\gamma \geq 1$, where $\gamma \to \infty$ signifies no disorder and $\gamma = 1$ is one order lesser than the hopping amplitude. Further, we assume a Gaussian-like form for $\sigma(x)$ centered at the junction and with a variance parameter $\sigma_D$,
\begin{equation}
    \sigma(x) \equiv \exp{\left[-\frac{1}{2}\left(\frac{x}{a\sigma_D}\right)^2\right]}
\end{equation}
where $a$ is the lattice discretization. This form of the variance profile of disorder is essentially a generalization of tunnel barriers at the interface, which accounts for random fluctuations as well as a decay of the fluctuations deep in the normal region. The variance parameter $\sigma_D$ thus denotes the extent of the disorder into the normal region. This is shown in Fig.~\ref{fig:figdisorder}(a). Thus, the random potential depends on the two parameters, $V(x) \equiv V(x;\gamma,\sigma_D)$, and the Hamiltonian is changed as follows:
\begin{equation}
    \mathcal{H} \mapsto \mathcal{H} + \sum_x V(x;\gamma,\sigma_D)c^{\dagger}_{x}c_x
\end{equation}
We take $L_n = 100$ lattice sites, which is computationally sufficient to study the effects of the disorder, which decays away from the junction. Each computation in Fig.~\ref{fig:figdisorder}(c) is averaged over $100$ disorder configurations. We have verified that averaging over higher number of configurations does not change the decay behavior of the pair amplitude. \\
\indent A schematic diagram of the correlation patterns in the system as a function of the decay rate $\gamma$ and the variance spread parameter $\sigma_D$ is shown in Fig.~\ref{fig:figdisorder}(b). The previous results in Sec.~\ref{sec:SNclean} were in the $\gamma \to \infty$ and $\sigma_D \to 0$ limit where we observed a power-law, or more generally algebraic decay in $1/r$ of the pair amplitude as a function of $r$. In the other limit of $\gamma \to 0$ and $\sigma_D \to \infty$, we expect the coherence properties of the normal region to be destroyed and thus infer a fully suppressed proximity-effect. We can indeed show, with limited computational resources on a finite size system, the transition between these two limits, as indicated in Fig.~\ref{fig:figdisorder}(b). In a more rigorous sense, we expect this to be a phase transition in the steady state $\rho_{\textsc{SS}}$ of the system, which we probe through the pair correlations in $\rho_{\textsc{SS}}$ using the methods developed before. The phase transition conjectured can be seen from the following simple argument upon defining the net induced order as $f \equiv \int_{x=0}^{L}F(x)dx/L$ for system size $L$. An algebraic decay of $F_c(x) \sim F_0\xi/(x+\xi)$ implies $f_c \sim (F_0\xi/L)\log{L}$ and exponential decay of $F_d(x) \sim F_0\exp{(-x/\xi)}$ implies $f_d \sim F_0\xi/L$ for large $L$ where subscripts $c(d)$ denote the  clean and disordered cases respectively. Clearly, $\lim_{L\to\infty}f_{c,d} = 0$ for both possibilities as the induced order decays to zero ultimately at $x\to\infty$. But, $f_c/f_d \sim \log{L}$, exhibiting a logarithmic jump at the critical disorder line. \\ 
\indent We plot in Fig.~\ref{fig:figdisorder}(c) the pair amplitude of four cases corresponding to variance constants $\sigma_D \in \{5,25,50,100\}$ and two values of the decay rate in each case to illustrate the crossover. In the case of $\sigma_D = 5$, keeping the decay very small $\gamma = 3$ we nearly recover the case of the clean nanowire as previously shown. When the decay is increased to $\gamma = 1$, we see a clear suppression in the induced pair amplitude. Nonetheless, upon zooming in (as shown in the inset) a power-law decay is seen. This indicates that the spread of disorder (parametrized by $\sigma_D$) in the normal region was low enough to still allow power-law correlations to survive even at $\gamma = 1$. Whereas, upon increasing the spread to $\sigma_D = 25$, we observe a power-law decay for $\gamma = 3$ but note an \textit{exponentially} suppressed pair amplitude in the normal region for $\gamma = 1$. This is shown in the two markers horizontally aligned with $-\gamma = -1$ in Fig.~\ref{fig:figdisorder}(b), with the boundary separating the two cases. Further increasing the spread of disorder into the normal region, at $\sigma_D = 50$ we see an exponential decay as early as at $\gamma = 1.75$, and at $\gamma = 2$ for the case of $\sigma_D = 100$. On the other hand, the decay for $\gamma = 3$ is power-law for all the variance constants. This in-turn informs the sketch drawn in Fig.~\ref{fig:figdisorder}(b). \\

\indent These results show that a systematic study of the effects of disorder is imperative in order to engineer proximity-induced correlations in real devices. Once a simplified model of the disorder profile in an device is estimated, the methods developed here can be used to analyze the experimental results in tandem with the expected behavior. In addition, it is also possible to include incoherent processes by including an effective self-energy that describes momentum and spin relaxation effects within the NEGF framework \cite{datta_dephasing,midha2024symmetry}.  
\subsection{Correlations in resonant Cooper-pair injectors}\label{sec:cpinj}
We will now demonstrate how our methods can be effectively applied to studying layered superconducting devices by analyzing the correlations in a device recently shown to exhibit enhanced Cooper pair injection via resonant tunneling \cite{CPInjection_ExpMain}. The transport properties of this device were previously studied theoretically using the Blonder-Tinkham-Klapwijk (BTK) model \cite{BTK} based on the BdG equations \cite{cpinjtheory}. A sketch of the device structure is outlined in Fig.~\ref{fig:figcpinj}(a), with the superconductor with characteristic gap $\Delta$ on the left side placed in proximity to a semiconducting region with an engineered potential creating a resonant quantum well. The essential physics of the device is well encapsulated in the creation of resonant levels between the semiconductor-superconductor barrier of height $H_2$ and the PN-junction barrier of height $H_1$ (see Fig.~\ref{fig:figcpinj}(a) for details on the parameters), allowing for controllable Cooper pair injection from the superconducting side into the normal region. The details of the device can be found in \cite{CPInjection_ExpMain,cpinjtheory}. \\
\indent In the absence of the superconducting region, the zero-bias conductance spectrum of the effective Fabry-per\'{o}t like cavity formed by the quantum well structure is shown in Fig.~\ref{fig:figcpinj}(b) for the chemical potential $\mu_{1} = 28.23$meV. This consists of regularly spaced peaks (resonances) of magnitude $\sim 2e^2/h$ as expected from a ballistic channel, which can be moved horizontally in the energy space by tuning the chemical potential. The proximity-induced correlations are expected to be sensitive to the modulation of transmission within the superconducting gap. Therefore, the chosen values for the quantum well parameters are designed to ensure that the width of the single peak in the zero-bias spectrum is narrower than the expected superconducting gap, thereby highlighting this effect. We now place the superconductor in proximity with an order parameter of $\Delta=1$meV, and plot the transmission of the hybrid device in Fig.~\ref{fig:figcpinj}(c) for $\mu = \mu_{1}$ as well as $\mu_{2} = 27.5$meV. One notes that the mid-gap conductance becomes greater than $\gtrsim 3.5e^2/h$ for $\mu = \mu_{1}$, owing to the presence of Andreev reflection at the junction resulting in a conductance greater than $2e^2/h$ and the fact that at $\mu=\mu_1$ there is a current peak inside the superconducting gap. We can tune out of this `resonant' state by turning the chemical potential to the `off-resonant' case of $\mu = \mu_{2}$, which removes the resonant sub-gap peak. This in-turn implies that the device can exhibit controllable Cooper pair injection from the superconductor into the semiconducting quantum well. \\
\indent We now analyze the pair amplitude throughout the device using the methods developed earlier, aiming to gain insight into the correlations generated within this device. The results are shown in Fig.~\ref{fig:figcpinj}(d). We observe that the pair amplitude inside the superconducting region (located to the left of the dotted vertical line) remains the same for both cases of $\mu \in \{\mu_{1},\mu_{2}\}$. However, we see a resonant increase in the pair amplitude inside the quantum well (normal) region for the case of $\mu_{1}$ as compared to that of $\mu_{2}$, in perfect correspondence with the conductance spectrum. We again stress the fact that the resonance is mirrored only inside the semiconducting region. Thus, the behavior of this Cooper pair injection device is fundamentally rooted in the underlying resonant mechanism of the pair amplitude inside the semiconductor. The transport measurements mirror this underlying resonance in the pair amplitude as an enhanced sub-gap conductance peak.\\ 
\indent We can further elucidate this by studying the spectral resolution of the induced correlations, as shown in Fig.~\ref{fig:figcpinj}(e)(i-v) for varying energy regimes $|E|/\Delta$ indicated on the plots. It is seen that the resonant behavior is rooted very deep inside the gap at energies $|E|/\Delta \leq 0.3$, in agreement with the previous discussions of Sec.~\ref{sec:spectral}. Further, as we move higher in energy, we note that a destructive interference between the energy-resolved correlations begins to take place for the off-resonant case of $\mu = \mu_{2}$, as seen by parity crossing oscillations in Fig.~\ref{fig:figcpinj}(e)(ii-iii). Whereas, for the resonant case of $\mu=\mu_1$, the energy resolved amplitudes do not take negative values inside the gap $|E| \leq \Delta$ as seen in Fig.~\ref{fig:figcpinj}(e)(i-iii). Moreover, this effect can still be seen slightly outside the gap in Fig.~\ref{fig:figcpinj}(e)(iv) for $1.1 \leq |E|/\Delta \leq 2$. Far outside, we note negligible contributions to the pair amplitude for both the cases with zero-mean oscillations Fig.~\ref{fig:figcpinj}(e)(v). Thus, it is crucially these spectral interference patterns which lead to the device functioning as a resonant Cooper pair injector. It is thus imperative to understand the nature of induced correlations to further engineer the injection efficiency and other merits of such devices. \\ 
\indent This discussion is broadly relevant to various devices designed to harness the superconducting proximity effect to create unique phenomena in non-ordered regions. Notably for the device discussed here, upon gaining such fundamental understanding of the mechanism enabling the resonance properties, one can systematically investigate the interference patterns in the energy-space to allow for optimization of injection efficiency in on-resonant scenarios. Moreover, a systematic examination of the induced correlations thus offers valuable insights into the fundamental mechanisms influencing transport properties, thereby paving the way for the development of advanced hybrid/layered quantum devices. 
\section{Conclusions}\label{sec:conc}
By examining prototypical one-dimensional superconducting-normal hybrids, we have developed a non-equilibrium Green's function technique to analyze proximity-induced superconducting correlations. Our techniques can be applied to a wide range of devices that use layered or stacked heterostructures and provide insight into microscopic effects at the layer interfaces which can be used to leverage the superconducting proximity effect for engineering various aspects of the quantum technology stack. \\
\indent We discussed the power-law decay of the induced pair amplitude in clean coherent nanowires. The effect of the Andreev approximation was then studied by analyzing the induced pair amplitude as well as through the spatial variation of the local density of states from the superconducting to the normal regions. The induced order parameter was then shown to have an intricate structure in the energy space, which was then used to gain new insight into a recently proposed resonant Cooper-pair injection device. Furthermore, we carefully investigated the impact of disorder on the induced correlations by introducing a random disorder potential inspired by experimental conditions. This led to the discovery of a disorder-induced crossover which manifested as a change in the correlation patterns, from power law to exponentially decaying in the normal region. It is thus interesting to further investigate the disorder-enhanced $p$-wave amplitude explored in \cite{disorderpwave} within our framework. \\
\indent Our method is applicable to higher-dimensional lattices and more complex superconducting couplings treated within the mean-field approximation, providing a general framework for understanding and proposing new experiments involving hybrid superconducting quantum devices. A key avenue for future research is to further explore the relationship between transport properties, disorder, and the induced correlations in these hybrid systems. For example, identifying the crossover threshold of disorder at which devices such as the Cooper pair injector cease to function as intended is an important task. This will require detailed analysis at finite temperatures, as well as under non-trivial voltage and thermal bias conditions.
\section{Acknowledgements}
The author BM wishes to acknowledge the support by the Science and Engineering Research Board (SERB), Government of India, Grant No. MTR/2021/000388. The authors SM and BM acknowledge support of the Dhananjay Joshi Foundation under an endowment to IIT Bombay. SM acknowledges fruitful discussions with Arnav Arora.
\bibliography{induced_bib}
\end{document}